\documentclass{article}
\usepackage[utf8]{inputenc}
\usepackage{times}
\usepackage{graphicx}
\usepackage{latexsym}
\usepackage[left=2.6cm, right=2.6cm, top=2.5cm, bottom=2cm]{geometry}
\usepackage[utf8]{inputenc}
\usepackage{amsmath}
\usepackage{amsthm}
\usepackage{algorithm}
\usepackage{multirow}
\usepackage{algpseudocode}
\usepackage{multicol}

\usepackage{xcolor}

\title{A Big Data Approach for Sequences Indexing on the Cloud via Burrows Wheeler Transform}
\author{Mario Randazzo\\DMI, University of Palermo \and Simona E. Rombo\\DMI, University of Palermo}
\date{}
\begin{document}
\maketitle
\begin{abstract}
  Indexing sequence data is important in the context of Precision Medicine, where large amounts of ``omics'' data have to be daily collected and analyzed in order to categorize patients and identify the most effective therapies. Here we propose an algorithm for the computation of Burrows Wheeler transform relying on Big Data technologies, i.e., Apache Spark and Hadoop. Our approach is the first that distributes the index computation and not only the input dataset, allowing to fully benefit of the available cloud resources.
\end{abstract}

\section{INTRODUCTION}
%The traditional page limit for ECAI long papers is {\bf 7 (six)} pages
%in the required format. The traditional page limit for short
%submissions is {\bf 2} pages.
%
%However, these page limits may change from one ECAI to
%another. Consult the most recent Call For Papers (CFP) for the most
%up-to-date page limits.

Precision Medicine aims to design individualized strategies for diagnostic or therapeutic decision-making, based on both genotypic and phenotypic information. It allows scientists and clinicians to understand which therapeutic and preventive approaches to a specific illness can work effectively in subgroups of patients based on their genetic make-up, lifestyle, and environmental factors \cite{Vassy287fs19}. The diffusion of high-throughput assays, such as next-generation sequencing (NGS) and mass spectrometry (MS), has led to fast accumulation of sequences and other omics data which can be used to enable Precision Medicine in practice. As an example, specific disease biomarkers may be identified by cleaning up raw data generated by NGS or MS, and then experimentally validated in laboratory. 

An important problem in this context is the indexing of NGS data \cite{ceri}. In particular, an index is a data structure that enables efficient retrieval of stored objects. Indexing strategies used in NGS allow  space-efficient storage of biological sequences in a  {\it full-text index} that enables fast querying, in order to return exact or approximate string matches. Popular full-text index data structures include variants of suffix arrays \cite{AbouelhodaKO04}, FM-index based on the Burrows–Wheeler transform (BWT) and some auxiliary tables \cite{FerraginaM05}, and hash tables \cite{pone}. The choice of a specific index structure is often a trade-off between query speed and memory consumption. For example, hash tables can be very fast but their memory footprint is sometimes prohibitive for large string collections \cite{schmit17}. 

Here we address the problem of computing BWT in the distributed, exploiting Big Data technologies such as Apache Spark \cite{spark}. In particular, previous research has been proposed on the BWT computation in a MapReduce \cite{DeanG10} fashion based on Apache Hadoop \cite{menon11}. The use of Spark and Hadoop together, as proposed here, have shown to notably improve the performance in several application contexts, due to the optimal exploitation of both memory and cloud. Another available tool relying on Hadoop and BWT computation is BigBWA \cite{DAbuinPPA15}. However, the BigBWA parallelism is intended only to split the input sequences and then apply another existing framework, i.e., BWA \cite{LiD09}, in order to align them via BWT. Therefore, the BWT computation is not based itself on Big Data technologies in BigBWA.

We propose an algorithm for BWT computation that fully exploits parallelism afforded by a cloud computing environment, combining advantages of MapReduce paradigm and Spark Resilient Distributed Datasets (RDD). The presented strategy is based on the computation of  Suffix Arrays in the distributed by revisiting the idea of \textit{prefix doubling} presented in \cite{FlickA15}. Validation results obtained on real biological datasets, including genomic and proteomic data, are provided, showing that our approach improves the performance for BWT computation with respect to its competitors. 

\section{METHODS}

\subsection{Background}

Let $S$ be a string of $n$ characters defined on the alphabet $\Sigma$. We denote by $S(i)$ the $i$-th character in $S$ and by $S_i$ its $i$-th suffix. We recall the following basic notions.

\paragraph{BWT} The Burrows-Wheeler transform of $S$ is useful in order to rearrange it into runs of similar characters. This may have advantages both for indexing and for compressing more efficiently $S$. The BWT applied to $S$ returns: 
\begin{itemize}
    \item a permutation $bwt(S)$ of $S$, obtained by sorting all its circular shifts in lexicographic order, and then extracting the last column;
    \item the index (0-based) $I$ of the row containing the original string $S$.
\end{itemize} 

\noindent Among the most important properties of BWT, it is reversible. Figure \ref{fig::bwt} shows an example of BWT for the string $S$=$BANANA\$$. In particular, $bwt(S)=BNN\$AAA$, and $I=3$.

\begin{figure}
    \centering
    \begin{tabular}{|c|c|}
    \hline
         {\bf All rotations of $S$} & {\bf Lexicographic sorting} \\
         \hline
         $BANANA\$$&  $ANANA\${\mathbf B}$\\
         $\$BANANA$& $ANA\$BA{\mathbf N}$\\
         $A\$BANAN$& $A\$BANA{\mathbf N}$\\
         $NA\$BANA$& $BANANA{\mathbf \$}$\\
         $ANA\$BAN$& $NANA\$B{\mathbf A}$\\
         $NANA\$BA$& $NA\$BAN{\mathbf A}$\\
         $ANANA\$B$& $\$BANAN{\mathbf A}$\\
         \hline
    \end{tabular}
    \caption{Example of BWT.}
    \label{fig::bwt}
\end{figure}

\paragraph{Suffix Array} 
The suffix array $SA$ of $S$ is defined as an array of integers providing the starting positions of suffixes of $S$ in lexicographical order. Therefore, an entry $SA[i]$ contains the starting position of the $i$-th suffix in $S$ among those in lexicographic order. Figure \ref{fig::sa} shows the Suffix Array for the same example of BWT.

\begin{figure}
    \centering
    \begin{tabular}{|c|l|l|c|}
    \hline
         {\bf i} & {\bf Suffixes} & {\bf Sorted Suffixes}&  {\bf SA[i]}\\
         \hline
         $1$& $BANANA\$$ &$ANANA\$$&$2$\\
         $2$& $ANANA\$$&$ANA\$$&$4$\\
         $3$& $NANA\$$&$A\$$&$6$\\
         $4$& $ANA\$$&$BANANA\$$&$1$\\
         $5$& $NA\$$&$NANA\$$&$3$\\
         $6$& $A\$$&$NA\$$&$5$\\
         $7$& $\$$&$\$$&$7$\\
         \hline
    \end{tabular}
    \caption{Example of Suffix Array.}
    \label{fig::sa}
\end{figure}

\paragraph{Inverse Suffix Array} 
Inverse Suffix Array of $S$, $ISA[i] = j$ means that the rank of the suffix $i$ is $j$, i.e., $SA[j] = i$.

\subsection{Proposed Approach}
The more crucial aspect for the BWT computation considered here is the calculation of the Suffix Array of the input string. Indeed, BWT can be calculated from the Suffix Array in a MapReduce fashion via join operation. Therefore, the algorithm proposed for the computation of the Suffix Array, based on the idea of \textit{prefix doubling} inspired by \cite{FlickA15}, is described below.     

\paragraph{Input:} 
Let $S$ be a string of length $n$, the input tuples set is:
$$\text{Input} = \{ (\texttt{null}, S(i)) : i = 1,\dots,n)\}$$
In the following, we assume that the input string $S$ ends with a sentinel character $\$$. This is typically used in the inverse transformation, however it is unimportant  for the purpose of the algorithm. 

\paragraph{Output:} A set of tuples of the form $(i, r)$, where $i$ is the index of a suffix in $S$ and $r$ is its rank (i.e., its position in the list of the sorted suffixes). In the literature this is referred to as the ISA. For our purpose, the resulting output is inverted in order to obtain the Suffix Array of $S$ and, then, its BWT.

\paragraph{Initialization: } The first step is starting from the Input set and initialize the set of tuples $(i, r)$, as described in the previous paragraph. In this phase, the rank is calculated by the first character of the suffix. In particular, let $Occ(c)$ be the number of occurrences of the character lexicographically smaller of $c$ in the string $S$, then the rank of the suffix $i$ can be determined as $Occ(S(i))$.\\
In a MapReduce fashion, this can be accomplished by first counting the occurrences of each character in $S$, and then computing the cumulative sum $Occ$ on the sorted counts. The map and reduce steps are:
$$\texttt{map: } (\texttt{null}, S(i)) \rightarrow (S(i), 1)$$
$$\texttt{reduce: } (c, \texttt{list}[1,1,\dots,1]) \rightarrow (c, \text{sum of ones}) $$
From this $Occ$ is calculated locally by collecting the result.\\
The $ISA$ set can be then initialized with the following map step:
$$\texttt{map: } (\texttt{null}, S(i)) \rightarrow (i, occ(S(i)))$$

\paragraph{ISA Extending:} The next step is to extend each rank contained in the initialized $ISA$ by the whole suffix. Here we use a technique called \textit{Prefix Doubling} that can be summarized as:
\begin{center}
	\textit{Given that the suffixes of a string are already sorted by their prefix of length $h$, we can deduce their ordering by their prefix of length $2h$.}
\end{center}
Given two suffixes $S_i$ and $S_j$ with an identical prefix of length $h$, we can deduce their sorting by comparing the order of the suffixes $S_{i+h}$ and $S_{j+h}$. Thus the idea is to pair, for each suffix $S_i$, its rank with the rank of the suffix $S_{i+h}$ (i.e., $(ISA[i]), ISA[i+h])$) and sort all these pairs in order to obtain the sorting by the prefix of length $2h$. Indeed, an iteration double the prefix, since the longest suffix has size $n$, all suffixes will be sorted after at most $\log_2(n)$ iterations.

\paragraph{Shifting and Pairing} To implement the above idea in a MapReduce fashion, we apply the two considered map steps to the latest $ISA$ calculated to obtain two different sets:
$$\texttt{map: } (i, r) \rightarrow (i, (r, 0))$$
$$\texttt{map: } (i, r) \rightarrow (i - 2^k, (-r, 0))$$
where $k$ is the number of the iterations minus one. 
The indices of rank are shifted this way, then the rank is paired by a reduce step. It is worth noticing that a negative number is used to denote a shifted rank, and the value is mapped as a tuple with a zero term in order to consider the ranks shifted that overflow the string length.\\
The union of the two obtained sets is considered and all tuples with a negative key are discarded (the corresponding ranks do not pair with any other rank in the set). The following reduce step is applied to the union:
$$\texttt{reduce: } (i, \texttt{list}[(r1,0), (r2,0)]) \rightarrow (i, (r1, -r2))$$
where $r2$ is the rank shifted. Some ranks may occur that are not reduced due to the unique key. These ranks overflow the length of $S$ and remain paired with zero. We denote the final set derived from this phase by $Pairs$.

\paragraph{Re-Ranking} Our purpose is to extend the previous rank with a new rank, obtained by considering the prefix doubled. Therefore, we compute the new rank according to the tuple in $Pairs$ as follows: firs we sort all tuples by value, then we compare each tuple at position $i$ (after sorting) with the one in position $i-1$. If they are equal, the new rank is equal to the rank of the previous tuple, otherwise the new rank is $i$. Finally, a new ISA set with rank extended is obtained, and the procedure is iterated on it again.
All operations described above can be achieved also in a distributed manner:
\begin{itemize}
    \item For the sorting operation, a certain number of partitions can be identified by range into roughly equal ranges the elements in the set (the ranges can be determined by sampling the data). Then for each partition a sorting algorithm is applied that sort each partition locally. This is easily provided by the framework Apache Spark. 
    \item In order to compute the new rank, the partition identified previously is considered and the procedure above is applied locally, as described before, using the length of the partition and the offset (i.e., the number of elements in the previous partition) for computing the position of the tuples.
\end{itemize}

\subsection{Example}
Let S = \textit{BANANA\$} be the input string of length $n = 7$. The input pairs are:  
\begin{equation*}
    \begin{gathered}
        \text{Input} = \{ (\texttt{null}, B), (\texttt{null}, A), (\texttt{null}, N),\\ (\texttt{null}, A), (\texttt{null}, N), (\texttt{null}, A), (\texttt{null}, \$) \}
    \end{gathered}
\end{equation*}
As for $Occ(c)$, it is shown in Table \ref{tab::occ}.

\begin{table}

\begin{center}
	\begin{tabular}{|c|c|c|c|c|}
		\hline 
		$c$ & $A$ & $B$ & $N$ & $\$$ \\ 
		\hline 
		$Occ(c)$ & 0 & 3 & 4 & 6  \\ 
		\hline 
	\end{tabular} 
\end{center}
    \caption{Computation of $Occ(c)$.}
    \label{tab::occ}
\end{table}

\noindent After the initialization, the initial ISA set is:
\begin{equation*}
    \begin{gathered}
        \text{ISA} = \{ (0, 3), (1, 0), (2, 4), (3, 0),\\ (4, 4), (5, 0),(6, 6)\}
    \end{gathered}
\end{equation*}
After the first iteration, the shifted tuples are:
\begin{equation*}
    \begin{gathered}
        \text{Shifted} = \{(-1, (-3, 0)), (0, (0, 0)), (1, (-4, 0)),\\ (2, (0, 0)), (3, (-4, 0)), (4, (0, 0)),(5, (-6, 0))\}
    \end{gathered}
\end{equation*}
After the the pairing we obtain the set:
\begin{equation*}
    \begin{gathered}
        \text{Pairs=}\{(0, (3, 0)), (1, (0, 4)), (2, (4, 0), (3, (0, 4),\\ (4, (4, 0), (5, (0, 6), (6, (6, 0)\} 
    \end{gathered}
\end{equation*}
Finally, we sort by value and we re-rank the indices. Then the new ISA is:
\begin{equation*}
    \begin{gathered}
        \text{ISA} = \{ (0, 3), (1, 1), (2, 4), (3, 1),\\ (4, 4), (5, 0),(6, 6)\}
    \end{gathered}
\end{equation*}
We observe that the only rank updated in this iteration is the one with index $5$, indeed shifting by $1$ it is possible to distinguish among the prefixes $AN$, $AN$ and $A\$$ corresponding to the suffixes $S_1$, $S_3$ and $S_5$.

\section{VALIDATION}
The presented algorithm has been evaluated on real datasets taken from the Pizza$\&$Chili website \cite{pizzachili}, where a set of text collections of various types and sizes are available to test experimentally compressed indexes. In particular, the text collections stored on this website have been selected to form a representative sample of different applications where indexed text searching might be useful. From this collection, we have chosen the following three datasets:

\begin{itemize}
    \item PROTEINS, containing a sequence of newline-separated protein sequences obtained from the Swissprot database. 
    \item DNA, a sequence of newline-separated gene DNA sequences obtained from files of the Gutenberg Project. 
    \item ENGLISH, the concatenation of English text files selected from collections of the Gutenberg Project. 
\end{itemize}

The main competitor of our approach is the algorithm proposed in \cite{menon11}, which is based on  MapReduce as well and divides the suffix array construction into multiple independent ranges that are then independently solved. In more detail, the partition points are selected and sorted, the
ranges are well-balanced across processing cores, and the final output forms a total order to the array.

We have implemented in Apache Spark both the algorithm described here and the approach proposed in \cite{menon11}, in order to provide a suitable comparison, and we have run them on the GARR Cloud Platform. In particular, we have configured the cluster with $1$ master and $48$ slave nodes, each node with $8$ VCore, $32$ GB of RAM and $200$ GB for disk. We have used Apache Hadoop $3.1.3$ and Spark $2.3.4$.

Results are shown in Table \ref{tab:time-result} (when the running time was larger than $10$ hours it has not been reported). For the PROTEINS dataset, it has been considered the only first $25$ MB, the only first $100$ MB and the full dataset. 

\begin{table}
\begin{center}
\begin{tabular}{|l|r|r|}
\hline
\multicolumn{1}{|c|}{\multirow{2}{*}{\textbf{Input}}} & \multicolumn{2}{c|}{\textbf{Time}}                                       \\ \cline{2-3} 
\multicolumn{1}{|c|}{}                                & \multicolumn{1}{l|}{Competitor} & \multicolumn{1}{l|}{Proposed Algorithm} \\ \hline
Proteins.200MB (25 MB)                                & 4.25 minutes                       & 3.86 minutes                        \\ \hline
Proteins.200MB (100 MB)                               & 3.05 hours                         & 11.36 minutes                       \\ \hline
Proteins.200MB (Full)                                 & -                                  & 22.03 minutes                       \\ \hline
Dna.200MB                                & -                                  & 26.78 minutes                       \\ \hline
English.1024MB                                        & -                                  & 3.4 Hours                           \\ \hline
\end{tabular}
\end{center}
\caption{Performance comparison between our algorithm and its competitor.}
\label{tab:time-result}
\end{table}

It is evident that our approach outperforms its competitor on all considered datasets. This is what we expected due to the fact that, as already discussed in the Introduction, we have presented the first method which introduces parallelism in the computation of BWT, allowing to fully benefit of cloud computing. It is worth notice that the algorithm in \cite{menon11} has been implemented here in Apache Spark, therefore it has been put in equal terms than our one in the comparison (i.e., the difference in performance cannot be referred to the fact that Spark allows optimizations with respect to Hadoop for a more efficient use of memory).

\section{CONCLUSION}
We have proposed a MapReduce algorithm for the implementation of a full-text index, that is, the Burrows Wheeler transform. We have implemented our approach in Apache Spark and we have proved by experimental evaluation that it is more efficient than its competitors already proposed in the literature. 

Among the various applications where an efficient and distributed implementation of BWT may be useful (e.g., data compression, pattern matching, etc.), and with a special attention to the Precision Medicine context, we mention that searching for a suitable combination of Indexing and Machine Learning techniques has recently proved to be a promising issue \cite{GRAHAM,raff2019new,FerraginaV20}. Therefore, we plan to focus our future studies in this direction.

\section{ACKNOWLEDGEMENTS}
Part of the research presented here has been funded by the MIUR-PRIN  research project ``Multicriteria Data Structures and Algorithms: from compressed  to learned indexes, and beyond'', grant n. 2017WR7SHH.

\bibliographystyle{plain}
\bibliography{ecai}

\end{document}